\documentclass[twocolumn,final]{elsarticle}
\usepackage{framed,multirow}
\usepackage{graphicx}
\usepackage{subcaption}

\usepackage{amssymb}
\usepackage{latexsym}

\usepackage{url}
\usepackage[pdftex,dvipsnames]{xcolor} % Coloured text etc.

\usepackage{xargs}                      % Use more than one optional parameter in a new commands
\usepackage[colorinlistoftodos,prependcaption]{todonotes}
\usepackage{hyperref}

\journal{Medical Image Analysis}

\begin{document}
%\verso{Nader Rafic \textit{et~al.}}
\begin{frontmatter}

\title{Using Deep Learning for an automatic detection and classification of the vascular bifurcations along the Circle of Willis}% \tnoteref{tnote1}}%

%\tnotetext[tnote1]{This is an example for title footnote coding.}

%\author[mymainaddress,mysecondaryaddress]{Anass Nouri }
%\author[mymainaddress1]{Romain Bourcier }
%\author[mymainaddress2]{Florent Autrusseau }
%\ead{Florent.Autrusseau@univ-nantes.fr}
%\address[mymainaddress]{SETIME Laboratory, Information Processing and Artificial Intelligence Team, Faculty of Sciences,
%Ibn Tofail University, BP 133, 14000 Kenitra, Morocco}
%\address[mysecondaryaddress]{ENSC, National School of Chemistry, Ibn Tofail University, BP 133, 14000 Kenitra, Morocco}
%\address[mymainaddress1]{Institut du Thorax, CHU de Nantes, Nantes Université, 44006, Nantes, France}
%\address[mymainaddress2]{LTEN \& RMeS lab, Nantes Université, Polytech’Nantes, 44306, Nantes, France}

\author[1]{Rafic Nader}
%\author[1]{Given-name2 \snm{Surname2}\fnref{fn1}}
%\fntext[fn1]{This is author footnote for second author.}
\author[1]{Romain Bourcier}
\author[1,2]{Florent Autrusseau\corref{cor1}}
\ead{Florent.Autrusseau@univ-nantes.fr}
\cortext[cor1]{Corresponding author: 
 Tel.: +33-240683156; }

\address[1]{Nantes Université, CHU Nantes, CNRS, INSERM, l’institut du thorax, F-44000 Nantes, France}
\address[2]{Nantes Université, Polytech’Nantes, LTeN, U-6607, Rue Ch. Pauc, 44306, Nantes, FRANCE}

%\received{XX February 2023}
%\finalform{XX April 2023}
%\accepted{XX April 2023}
%\availableonline{XX May 2023}
%\communicated{---------}

\begin{abstract}
%%%
Most of the intracranial aneurysms (ICA) occur on a specific portion of the cerebral vascular tree named the Circle of Willis (CoW). 
More particularly, they mainly arise onto fifteen of the major arterial bifurcations constituting this circular structure. 
Hence, for an efficient and timely diagnosis it is critical to develop some methods being able to accurately recognize each Bifurcation of Interest (BoI).
Indeed, an automatic extraction of the bifurcations presenting the higher risk of developing an ICA would offer the neuroradiologists a quick glance at the most alarming areas.
Due to the recent efforts on Artificial Intelligence, Deep Learning turned out to be the best performing technology for many pattern recognition tasks. Moreover, various methods have been particularly designed for medical image analysis purposes.
This study intends to assist the neuroradiologists to promptly locate any bifurcation presenting a high risk of ICA occurrence. 
It can be seen as a Computer Aided Diagnosis scheme, where the Artificial Intelligence facilitates the access to the regions of interest within the MRI. 
In this work, we propose a method for a fully automatic detection and recognition of the bifurcations of interest forming the Circle of Willis. 
Several neural networks architectures have been tested, and we thoroughly evaluate the bifurcation recognition rate.
%%%
\end{abstract}

\end{frontmatter}

%\linenumbers


%%%%%%%%%%%%%%%%%%%%%%%%%%%%%%%%%%%%%%
\section{Introduction}

A cerebral aneurysm is a bulge or dilation of an artery in the brain resulting from a weakness in the blood vessel wall \citep{brisman2006,chalouhi2013}. Untreated brain aneurysms may present a risk of rupture, leading to an hemorrhagic stroke potentially causing the patient's death for up to 50 $\%$ of all cases \citep{frosen2012,suarez2006,van2001}.
MRI or CT scans are generally used to detect the presence of intracranial aneurysms \citep{Jerman2015,CADcontrib,Duan2019,Dai2020,Sichtermann2019,Shahzad2020,Stafa2008,Timmins2021,Ueda2019,yang2021}. Such lesions occur most frequently along the Circle of Willis (CoW), onto a particular set of arterial bifurcations \citep{BROWN2014,keedy2006}. The Circle of Willis consists of a set of arteries in the base of the brain that connects the left and the right anterior cerebral trees to the posterior cerebral tree \citep{LECLERC2003}. There are many different CoW configurations, with some variations in the number, shape and size of arteries \citep{csahin2018,hartkamp2000,jones2021}. As a consequence, these variations affect the blood flow within the brain and may be associated with several cerebrovascular pathologies \citep{pascalau2019}. Understanding the geometric variations of the arteries and bifurcations is important to determine risk factors for numerous vascular problems including aneurysm occurrence \citep{bogunovic2013,kayembe1984,lazzaro2012}. 
Due to the increasing workload and the demanding nature of the detection process by radiologists, an automatic detection and recognition of the bifurcations of the CoW can be a great help to establish a diagnosis, to detect and to monitor aneurysms at an early stage. 
An automatic recognition of the bifurcations of interest could be used jointly with vascular tree characterization methods, in order to anticipate an ICA formation, due to an abnormal bifurcation configuration. 
Indeed, in recent works~\citep{Nouri2020}, we have proposed some tools for a full geometric characterization of the CoW bifurcations. However, this study was conducted on the MRA-TOF images acquired from 25 patients presenting an aneurysm either on the left or the right Middle Cerebral Artery (MCA). In this previous investigation, we did limit our dataset to 25 images only, as the bifurcations were manually selected, and a full evaluation of the collected features was quite a tedious process. Moreover, we only focused onto MCA aneurysms to reduce a little the annotation burden. Obviously, an automatic bifurcation recognition process would have been extremely useful to expand the study to a much larger dataset.
Thanks to the method proposed in this paper, such an analysis could be conducted again with a much larger dataset and onto any other bifurcation of Interest of the Circle of Willis, hence guiding the neuroradiologists toward a better understanding on the risk of ICA formation (or even the risk of evolution). 

Indeed, we have recently showed that a full bifurcation characterization may also prove very useful to monitor the bifurcation evolution during the aneurysm growth~\citep{Boucherit2021}. 

In order to diagnose various brain vessel deformities or cerebrovascular pathologies, brain vascular segmentation approaches have been widely studied in the literature \citep{MOCCIA2018,klepaczko2016,cetin2015,rempfler2015,xu2010}.
However, fewer works have proposed cerebral bifurcation detection and labeling \citep{bogunovic2011,bogunovic2013,robben2013,Robben2016,gurobi2015,bilgel2013,dunaas2017}. The bifurcations correspond to the endpoints of arteries and more precisely the intersection between merging or splitting branches. Thus, classifying the CoW bifurcations allows to identify the connecting arteries. 
Using Maximium a Posteriori probability estimation, Bogunovic \textit{et al.} \citep{bogunovic2013} automatically labeled $11$ of the main bifurcations of the CoW based on their attributes, their topology and their probability of occurrence and they obtained high accuracy results (95$\%$) however this score drops down to $58\%$ when the entire CoW is under consideration. 
In \citep{dunaas2017}, Dunas \textit{et al.} created an automated pipeline based on a probabilistic atlas, describing the location and the shape of the main brain arteries. 
Robben \textit{et al.} \citep{Robben2016} proposed a method that simultaneously segments and labels the cerebral vasculature by a probabilistic formulation. The latter produces an integer optimization program solved using the branch-and-cut algorithm \citep{gurobi2015}.
In \citep{bilgel2013}, the authors applied a Random Forest classifier to the features computed on the vessel centerlines of the cerebral arteries, thus obtaining a class probabilities for each blood vessel segment. 
These likelihoods are then used as input of a Bayesian Belief network for a better classification accuracy. 
Wang \textit{et al.}~\citep{wang2017} used a supervised machine learning algorithm and more specifically XGBoost classifier for the anatomical labeling of $11$ bifurcations belonging to the Circle of Willis. To classify the bifurcations, they used several geometrical features defined on a vessel centerline model. The probabilities obtained by the classifier are then used as inputs to a Hidden Markov Model (HMM) to impose some topology constraints for a more accurate labeling. 
Finally in \citep{essadik2022}, the authors also considered Machine Learning methods to identify the main bifurcations of the cerebral arteries by using various geometric features \citep{Bifurcpatent2018} of the cerebral arteries as input of supervised ML algorithms. 
They obtained a rather efficient labeling of the main cerebral artery bifurcations. However, all these methods need a first pre-processing step isolating CoW arteries before resorting to the clustering step, which limits the use of such approaches in a real world scenario. 
The initial bifurcation labeling is commonly performed by highly trained, experienced readers.
Recently, Convolutional Neural Networks (CNNs) have become the standard approach in medical image analysis achieving state-of-the-art results in various tasks, such as classification, object detection and segmentation \citep{YU202192,pouyanfar2018,liu2017}. In this context, CNNs have shown promising results on brain vessel segmentation \citep{guo2021,meng2020,fan2020,zhang2020,livne2019,hilbert2020,sanchesa2019,Tetteh2020}. 
\begin{figure*}[!ht]
\begin{centering}
\includegraphics[width=.9\linewidth]{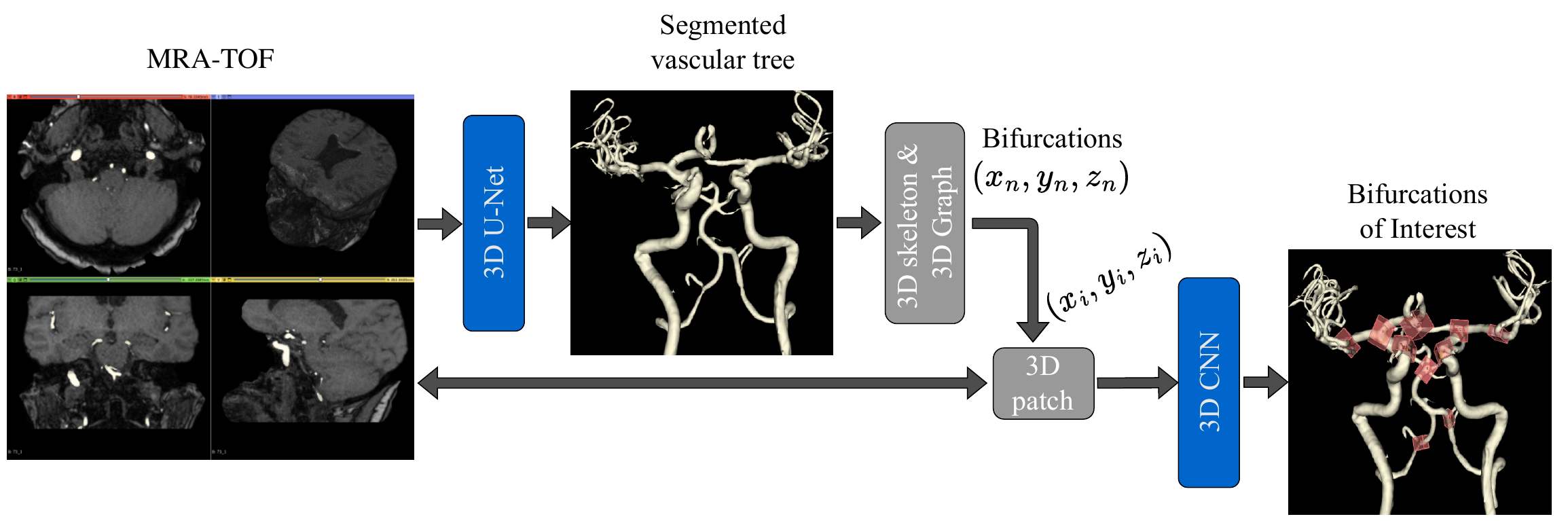}
\par\end{centering}
\caption{General flowchart of the bifurcation recognition process.\label{GlobalScheme}}
\end{figure*}
In \citep{DUMAIS2022}, the authors created an automated deep learning based method to segment and label the main arteries composing the CoW. Although our work uses a similar approach, we don't tackle exactly the same issue. In this study, we propose a classification method for automatic labeling of the main CoW bifurcations responsible for about $85\%$ of the cerebral aneurysms occurrences. The labeling of these intersecting points between arteries extends the work on CoW arteries segmentation considered in \citep{DUMAIS2022}.
The work presented in this paper is a part of a large series of scientific research projects (National French projects \textit{ANR-ICAN}, \textit{ANR-WECAN} and \textit{PHRC-UCAN})~\citep{Bourcier2017,Neurosurg2020,Boucherit2021}, where neuroradiologists aim to identify and understand the multiple factors that lead to the development of saccular aneurysms along the Circle of Willis. Notably, they intend to fully characterize the geometrical features of the bifurcations, so as to estimate the risk of ICA occurrence depending on a given bifurcation configuration.
These interdisciplinary projects intend to study the ICA formation from both the genetic predisposition and image analysis perspective.
Thus, a significant part of the project is dedicated to medical image analysis. Part of our investigations is devoted to the automatic detection and the characterization of the CoW bifurcations \citep{Nouri2020,essadik2022},
synthetic generation of brain vasculature \citep{autrusseau2022,chater2021}, and the automatic detection and characterization of aneurysms in Magnetic Resonance Angiography (MRA) - Time of Flight (TOF) acquisitions (ongoing works).
In this study, we propose an automated detection and labeling of the major CoW bifurcations using deep learning-based methods. 

The rest of the paper is organized as follows. Section 2 describes the pipeline of the proposed method including the considered dataset, the neural networks architectures and the evaluation metrics. In Section 3, we present the experimental results. We provide a thorough analysis of the method's performances, we compare different CNN architectures, and try to understand the variability of the recognition rates per bifurcation. Finally, Section 4 provides a discussion on the proposed method and concludes this work.

%%%%%%%%%%%%%%%%%%%%%%%%%%%%%%%%%%%%%%
\section{Materials and Methods}

In this section, we will thoroughly describe the image dataset that has been constituted, the annotation process, the CNN architecture we have been using, as well as a proposed U-Net based segmentation method, needed to come up with a fully automatic process.
For a given test image, we show in Fig.~\ref{GlobalScheme} the global flowchart of the bifurcation recognition process. We can observe how the two distinct neural networks are being successively used; the first one (3D U-Net) provides a vasculature segmentation, whereas the second (3D CNN) is applied onto 3D patches, centered around the vascular bifurcations for BoI classification. 
In this section, we will describe in details each and every step of this process. 
\begin{table*}[!t]
\caption{ Summary of the Time of flight (TOF) magnetic resonance imaging (MRI) dataset used in the study. \label{tab1}}
\centering
\begin{tabular}{|c|c|c|c|c|}
\hline
\textbf{Dataset}&\textbf{Constructor}&\textbf{MR device}&\textbf{MFS (T)}&\textbf{Participants}\\
\hline
Set-1&GE Medical Systems & Optima MR450W&1.5&25\\

\hline
Set-2 & GE Medical Systems & Discovery MR750W&3.0&9\\

\hline
Set-3 & GE Medical Systems & Signa HDxt&1.5&1\\

\hline
Set-4&SIEMENS&Aera&1.5&32\\

\hline
Set-5 & SIEMENS & Skyra & 3.0 & 20\\

\hline
Set-6&SIEMENS & Avanto & 1.5 & 3\\

\hline
Set-7 & SIEMENS & Prisma & 3.0 & 3\\

\hline
Set-8 & SIEMENS & Sonata & 1.5 & 3\\

\hline
Set-9 & SIEMENS &Verio & 3.0 & 1 \\

\hline
Set-10 & Philips Medical Systems& Ingenia&3.0&19 \\
\hline
Set-11& Philips Medical Systems& Achieva &3.0&17\\
\hline
Set-12& Philips Medical Systems & Achieva&1.5&2\\
\hline
\end{tabular}
\end{table*}
%%%%%%%%%%%%%%%%%%%%%%%%

\subsection{Data acquisition}

In this work, we have gathered an heterogeneous MRA dataset (148 patients) from the ICAN database. This database consists of MRA-TOF images acquired on a large panel of patients (more than $2\,500$) of various ages, within about 60 different medical institutions all over the French territory. In the current study, among the $2\,500$ images constituting the ICAN dataset, 148 images were randomly chosen. These images were issued from 28 French institutions and hence, acquired on various MR machines with a wide range of MR acquisition parameters, thus ensuring the robustness of a deep learning model that could generalize across different datasets. The MRA-TOF images were acquired on 12 different MRI scanners (from Siemens Healthcare, GE Medical systems and Philips Medical systems) partitioning the data into 12 different sets. The acquisition parameters are reported in Table \ref{tab1}. Overall, 118 MRA-TOFs were used for training and 30 images were retained as an independent test dataset.  To guarantee that all MRA-TOFs have coherent dimensions and voxel spacing, the images were re-sampled to a median voxel spacing of $0.4{mm}^{3}$ \citep{isensee2021}.

\subsection{Data Preparation/Annotation}
Our main goal is to detect the bifurcations of interest along the Circle of Willis in 3D medical volumes using Convolutional Neural Networks (CNNs) which is among the most promising solutions in medical image analysis due to its high capability in extracting powerful high-level features. 
In order to fully exploit the spatial contextual information, we have opted for 3D CNNs, \textit{i.e.} based on 3D convolution operations, rather than sequentially employing a 2D CNN on the MRA slices. Indeed, this commonly ensures a better detection \citep{singh2020}.
However, applying 3D CNNs is not straightforward. Processing the entire brain volume is very challenging, it requires significant GPU memory and is computationally expensive and time-consuming. To tackle these problems, we have redesigned the 3D bifurcation recognition as a patch-wise classification task, the network was thus trained on 3D patches encompassing the bifurcations of interest of the CoW. It is thus necessary to precisely locate each and every BoI within the 148 TOFs. This  was possible by a pre-segmentation step. For this purpose, two distinct annotation processes had to be manually performed : \textit{i)} a full vascular tree segmentation, and \textit{ii)} position markers had to be placed onto each BoI. 
An initial annotation was first conducted by a trained operator (author FA), and a medical expert (author RB : neuroradiologist with 10 years experience) validated and corrected whenever necessary each and every labeled image of the dataset.

\subsubsection{Step 1-Centerline extraction}
Thanks to the vasculature skeletonization, we can locate all the bifurcations centers within the 3D space. 
The approach described in \citep{Nouri2020} was used. But first, for each image, the cerebral arteries must be  properly segmented. The vascular tree segmentation can be obtained using various methods as reviewed in \citep{MOCCIA2018}. 
In order for the bifurcation recognition method to be fully automatic, an efficient vascular tree segmentation algorithm must be applied. We will present in section~\ref{sc} the Deep Learning-based segmentation method (U-Net) we have considered. 
Thus, for the training phase of the segmentation network, the annotations were manually performed by a trained operator using the 3D Slicer software\footnote{https://www.slicer.org}. 
Given the 3D cerebral vasculature segmentation,  we compute its skeleton using the method described in  \citep{lee1994}, then, a 3D undirected graph \citep{Bifurcpatent2018} is generated from the extracted skeleton. The graph edges represent the centerlines of the arteries and the nodes represent either the bifurcations or the loose ends of the arteries. 

\begin{figure}[!ht]
\begin{centering}
\includegraphics[width=.8\columnwidth]{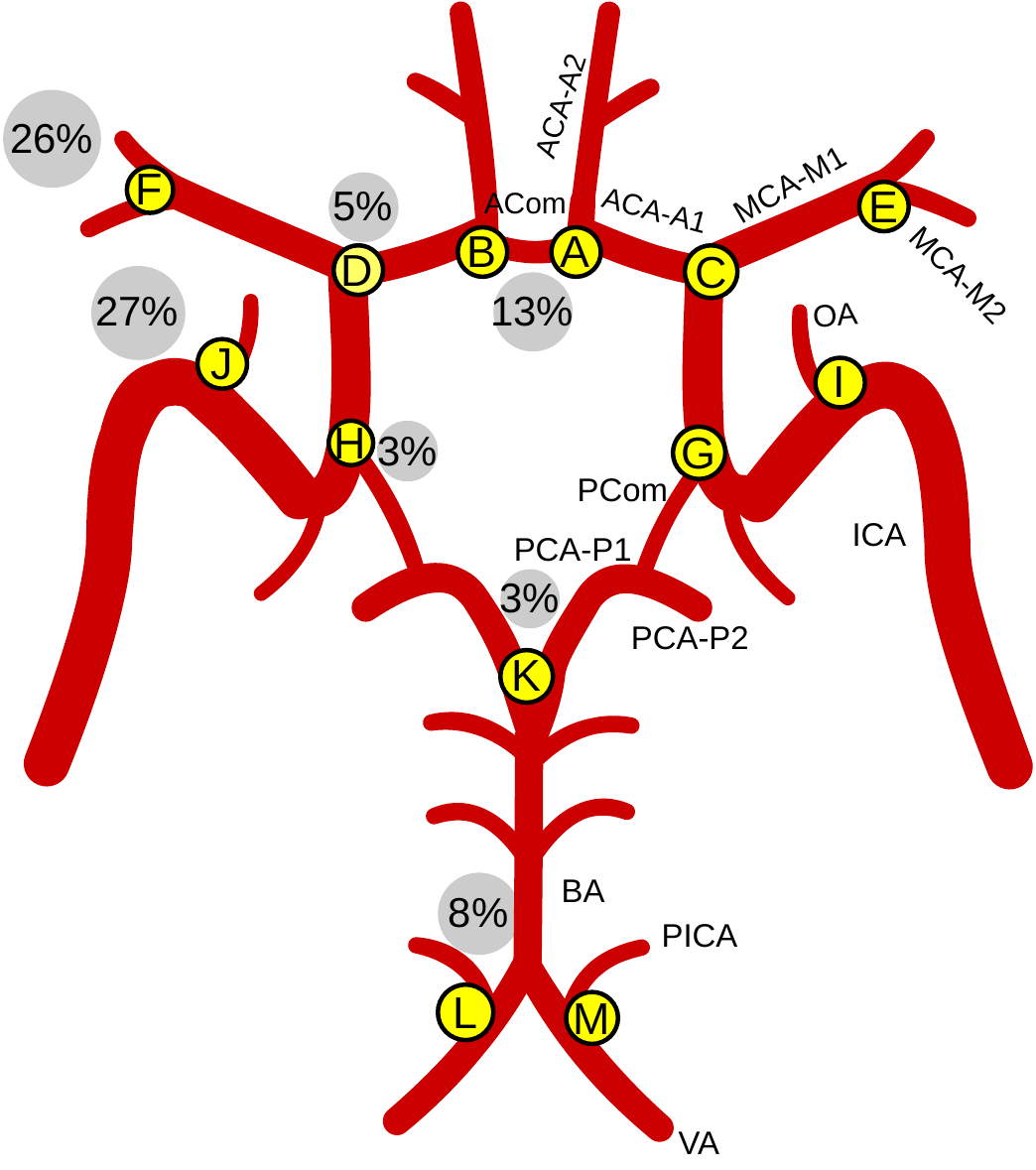}
\par\end{centering}
\caption{Schematic representation of the Circle of Willis; Letters in yellow discs stand for the bifurcations labels, whereas the percentages within the gray discs show the frequency of aneurysms occurrence.\label{willis}}
\end{figure}

\subsubsection{Step 2-Patch selection}

Using the 3D graph, one can easily locate the center of any vascular bifurcation in the 3D space. Hence, using the manually positioned labels, we have located all the CoW bifurcations and cropped a surrounding patch ($32 \times 32 \times 32$ voxels) around the center coordinates. Based on our preliminary tests, these 32 voxels patches adequately encompassed each bifurcation of interest while preventing to crop two consecutive bifurcations of interest within the same patch.

In this study, we have selected 13 BoIs representing the highest risk of aneurysm occurrence \citep{Robben2016} described as follows (bifurcations labeled A to M in Fig.~\ref{willis}).
The intersection points between the first and the second segments of the anterior cerebral artery \textit{ACA-A1} and \textit{ACA-A2}, for the left and right anterior cerebral tree (A and B), the bifurcations between \textit{ACA-A1} and the middle cerebral arteries \textit{MCA} (C and D), the division points of the \textit{MCA} into the segments \textit{M1} and \textit{M2} (E and F). Other sites include the internal carotid artery (G and H) and ophthalmic arteries \textit{OA} (I and J). In the posterior cerebral tree, the common locations where aneurysm development is high include the intersection of the basilar artery \textit{BA} and the posterior communicating arteries (K) and finally the 2 sites spanning from the vertebral arteries to the posterior inferior cerebral arteries (M and L). Overall, this makes 13 bifurcations of interest for the classification task. 

A $14^{th}$ class is devoted to all the remaining bifurcations (Bifurcations of Non-Interest). These bifurcations will be referred to as BoNI from now on. The purpose of this $14^{th}$ category is to discern the CoW bifurcations from all others, being outside the CoW. This allows an automatic detection to be applied on the entire MRA-TOF of a patient. 
It is worth noting that the constituted dataset is unfortunately not composed of a balanced amount of each bifurcation of interest. This is due to the large variability of CoW configurations among patients, explained by missing, duplicated or hypoplastic arteries \citep{csahin2018,hartkamp2000,jones2021}. The variability could also be caused by the MR acquisitions where in certain cases, the posterior cerebral tree is not complete. More details on the distribution of classes in our dataset can be found in Table \ref{tab2}.
\begin{table}
\caption{Number of patches for each bifurcation category in the data set. \label{tab2}}
\begin{center} 
\begin{tabular}{|c|c|}
\hline
\textbf{Bifurcation label}&\textbf{Data set (118 TOFs)}\\
\hline
$\#$BoNI & 212\\
\hline
$\#$A &81\\
\hline
$\#$B &73\\
\hline
$\#$C &93\\
\hline
$\#$D &92\\
\hline
$\#$E &100\\
\hline
$\#$F &95\\
\hline
$\#$G &59\\
\hline
$\#$H &55\\
\hline
$\#$I &70\\
\hline
$\#$J &81\\
\hline
$\#$K &94\\
\hline
$\#$L &23\\
\hline
$\#$M &25\\
\hline
\end{tabular}
\end{center}
\end{table}

\subsection{Deep Learning-based segmentation of brain vessels}\label{sc}

For a fully automated recognition of the BoIs, some candidate patches must be first extracted using the segmented image. The segmentation is performed by a U-Net architecture \citep{ciccek2016,ronneberger2015}, which presents an encoder/decoder structure. 
Each level in the encoder consists of two successive 3D convolution layers (kernel size: $3 \times 3 \times 3$ , stride: 1), a batch normalization layer and a Rectified Linear Unit (ReLU) activation, followed by a max pooling layer (kernel size: $2 \times 2 \times 2$, stride: 1). The depth of the feature maps doubles with each downsampling (going from 32 to 256). For the decoder path, max pooling layers were replaced by up-sampling layers. Furthermore, training is performed using Adam optimizer with a learning rate of 0.0001, a batch size of 8 and the Dice loss as a cost function. We use a threshold of 0.5 to binarize the  U-Net outputs. 
The segmentation task is applied on 3D patches of size $64 \times 64 \times 64$ extracted from each MRA-TOF. In total, 100 patches were extracted per TOF, among those, 70 patches were centered onto a blood vessel, whereas the remaining 30 were not.
From 118 MRA-TOFs, 88 were used for training and 30 for validation. The U-Net performance was measured using the whole-brain segmentation constructed from aggregating patch-wise predictions and  was conducted on the test dataset using  four different metrics : Precision and Recall to assess the quality and the completeness of the segmentation, Dice similarity Coefficient (DSC) to quantify the overlap between the ground truth and the prediction, and 95 percentile Hausdorff Distance (95HD) to capture the boundary errors. 
%%%
 The primary goal in developing a reliable bifurcation recognition method is not solely focused on reaching optimized segmentation performances. Instead, the emphasis lies on effectively extracting accurate bifurcation locations from the segmented image and the subsequent 3D skeleton representation.

%%%
Thus, in order to evaluate the fidelity of the 3D graph issued from the skeletonized segmentation  along the Circle of Willis, we have performed a structural assessment of the extracted bifurcations. That is to say, we evaluate the presence/absence of the various arteries constituting the CoW. This was achieved by overlaying the predicted output masks onto the original scans, and having the two mentioned trained annotators visually assess which arteries were correctly identified for each patient based on a predefined scheme of Circle of Willis arteries. 
Each single artery was considered as being properly identified if at least a portion of the segmented branch emerges from the bifurcation at the graph node and overlaps with the artery as observed on the raw MRA scan.

%%%%%%%%%%%%%%%%%%%%%%%%
\subsection{Classification networks architectures}

\begin{figure*}[!ht]
\begin{centering}
\includegraphics[width=0.9\linewidth]{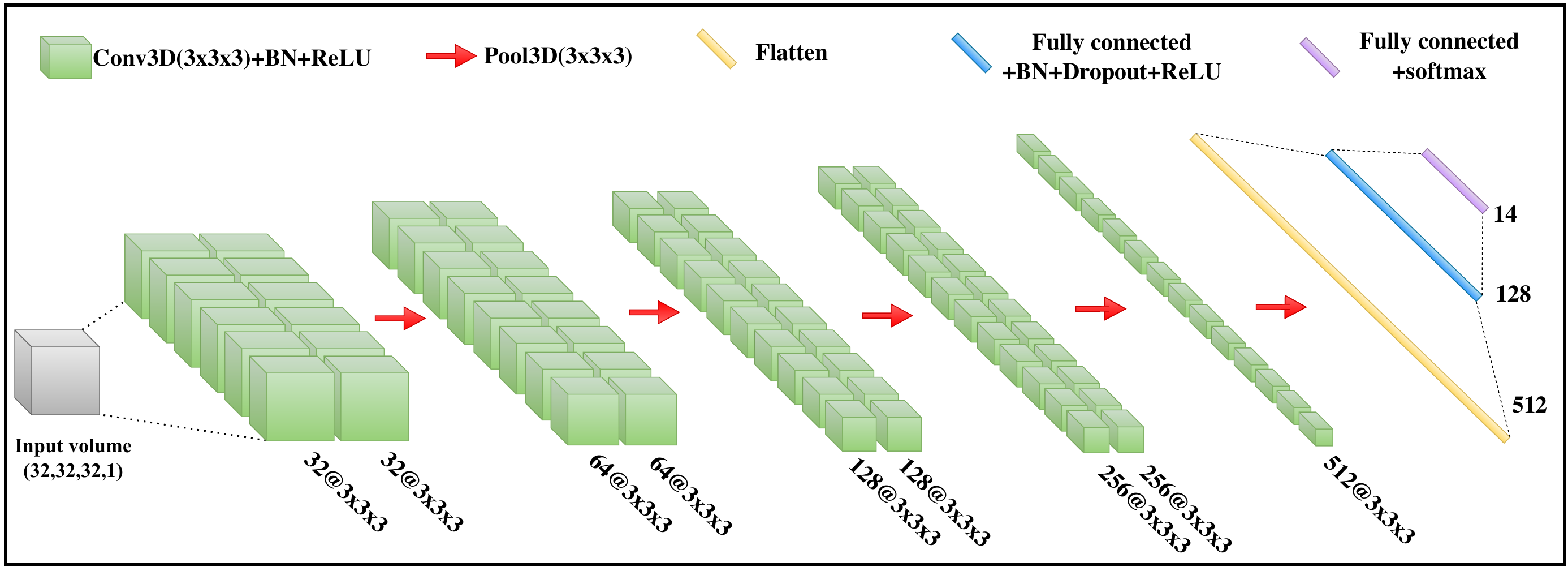}
\par\end{centering}
\caption{ Architecture of our 3D CNN. \label{cnn_arch}}
\end{figure*}

With regard to medical image classification, a few research groups designed their own networks from scratch while most used established architectures (VGG, ResNet, DenseNet) that achieve state-of-the-art performances \citep{rawat2017deep} on very large public datasets of 2D images (Imagenet, Coco, etc.). However, when it comes to medical image classification with little data, CNN with few layers could achieve, with a shorter training time, comparable or even better performances than deeper and more complex architectures. To the best of our knowledge, no research project has ever been conducted on the classification of the CoW bifurcations using Deep Learning. However, we do not intend to run a full performance benchmark of the state of the art methods applied for our particular task. We have conducted a study considering three widely used image classification models, VGGNet \citep{sim15}, ResNet \citep{he2016}, and DenseNet \citep{huang2017}.

In order to preserve the information along the depth dimension, we used 3D CNN networks obtained through 3D convolution layers. As previously described, we considered small $32\times32\times32$ patches as the network's input. All patches were z-score normalized. Considering this patch size, and a CNN with a final softmax layer, giving the probabilities for each of the 14 classes, we have considered different  architectures. We have first re-designed the three mentioned CNN networks (VGG-16, ResNet-18 and DenseNet-121) by replacing the 2D layers by their respective 3D layers while keeping the same number of layers. It is worth mentionning that for each model, we chose the version with the smallest number of layers (16 for VGG, 18 for ResNet and 121 for DenseNet). Moreover, we have  conceived a 3D CNN inspired by the VGG-16 model by keeping 9 convolutional layers (kernel size: $3 \times 3 \times 3$, stride: 1) grouped in 5 convolutional blocks with 32, 64, 128, 256 and 512 feature channels respectively. In order to reduce the input dimensionality, max pooling layers (kernel size: $2 \times 2 \times 2$, stride: 2) were considered throughout the network. 
Every convolutional layer is followed by a 3D batch-normalization layer (BN) and a rectified linear unit (ReLU) activation layer. The last 2 layers are fully connected.
The architecture of our 3D CNN is depicted in Fig.~\ref{cnn_arch}. Due to the small number of training samples, we kept the network relatively small, thus avoiding overfitting. Finally, a dropout layer with a $60\%$ rate is used for regularization before the final layer of all the considered networks.

%%%%%%%%%%%%%%%%%%%%%%%%
\subsection{Training}

All previously described CNNs were implemented using Tensorflow. Random flipping with a probability of 0.5 around the z-axis (number of slices) was performed to increase the training set. Other transformations have not shown any improvement in the classification accuracy. Axial flipping was carefully done changing the bifurcation label to its corresponding new value (left anterior bifurcation label turned into the right label and vice-versa).
Convolution layers were initialized using Xavier initialization \citep{glorot2010}. Optimization was performed using the categorical cross-entropy loss, Adam optimizer \citep{kingma2015} with $\beta_1=0$ and $\beta_2=0.9$, and a learning rate of 0.0001. All models were trained for 250 epochs with a batch size of 32.

%%%%%%%%%%%%%%%%%%%%%%%%
\subsection{Evaluation and model selection}

\begin{figure}[!ht]
\centering
\includegraphics[width=.8\columnwidth]{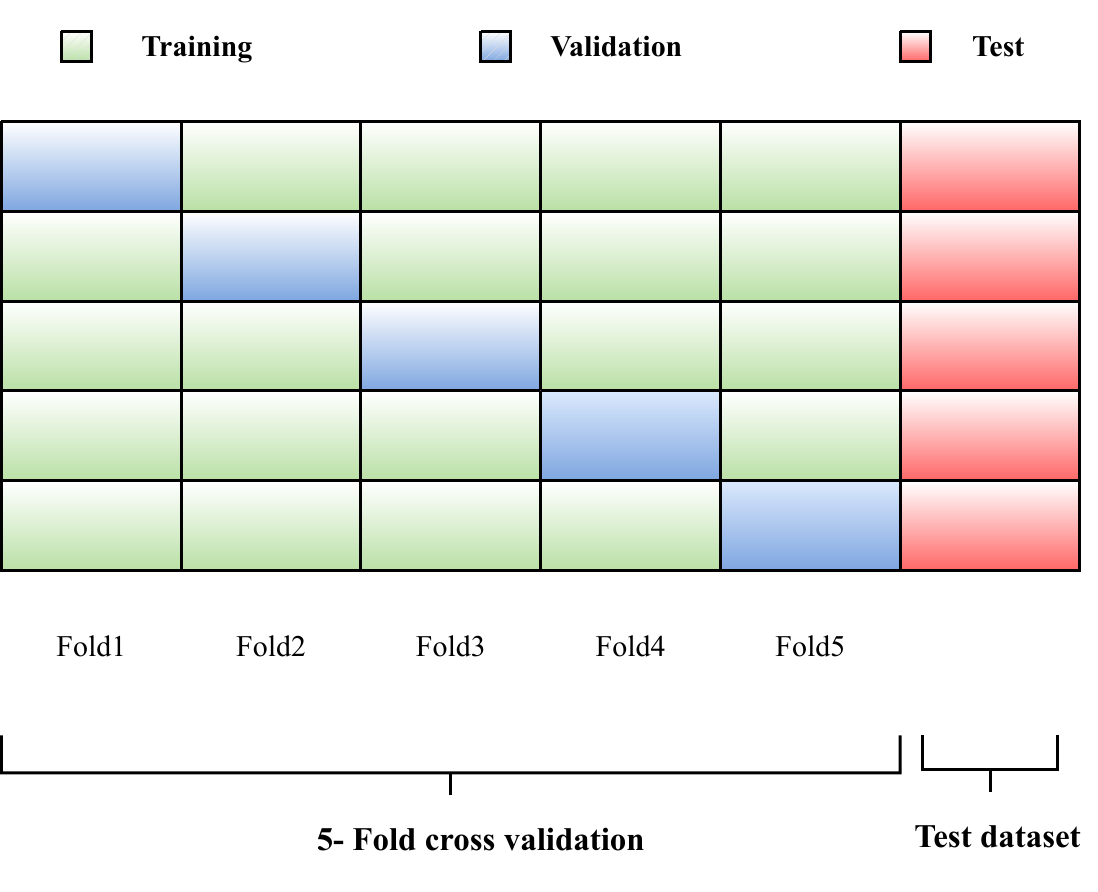}
\caption{Data splitting and cross-validation framework. Each fold comprises around 24 images and the separate test dataset is composed of 30 images. \label{crossValSplit}}
\end{figure}

A five-fold cross-validation with respect to the patients was used for the evaluation of each model. For this purpose, the training set (118 MRA-TOFs) was split into 5 distinct subsets making sure that each one has a similar distribution of the different classes (bifurcations of interest). As a result, five models will thus be trained, each with 4 of the folds being used as the training data and the remaining fold as validation data. Figure \ref{crossValSplit} illustrates the data splitting and cross-validation procedure.

To compare the different models, we have computed the area under the ROC Curve (AUC). The ROC curve is plotted with the True Positive Rate (TPR) against the False Positive Rate (FPR) at different thresholds, where TPR and FPR are defined as:

$$TPR=\frac{TP}{TP+FN},$$

and,

$$FPR=\frac{FP}{FP+TN}.$$

The AUC measures the entire two-dimensional area underneath the ROC curve. In parallel, we also compute the accuracy score and F1-score defined as follows:

$$ACC=\frac{TP+TN}{TP+TN+FP+FN},$$

and,

$$F1\mbox{-}score=\frac{2TP}{2TP+FP+FN}.$$

For all the validation samples and for each of the folds, the above metrics were computed, thus providing an evaluation of the whole dataset.
The holdout testing set was based on these five-fold models. The so-obtained predictions were then averaged from the five models to calculate the final metrics on the testing set.

%%%%%%%%%%%%%%%%%%%%%%%%
\subsection{Patient-wise evaluation}

The cerebral vasculature is composed of the CoW arteries but also of other brain distal vessels. As the 3D undirected graph is carried out on the whole segmented vascular tree, hundreds of bifurcation patches are extracted per image. The vast majority correspond to the BoNI (\textit{i.e.} outside the Circle of Willis), while the remaining patches are those surrounding the BoIs. These latter are centered around the 3D graph vertices. 
Let us recall that for training, we have applied a selection of  patches based on the location of a particular BoI in the MRA-TOF and its approximate distance to a node of the 3D graph.
Hence, a given BoI center may not correspond to a unique node of the 3D graph but will likely be located in its vicinity. 
This could partially be explained by some imperfections due to either the skeletonization or the segmentation process, but also, by a significant well known variability of the cerebral vasculature anatomy. % }

\begin{figure}[!t]
\centering
\includegraphics[scale=.3]{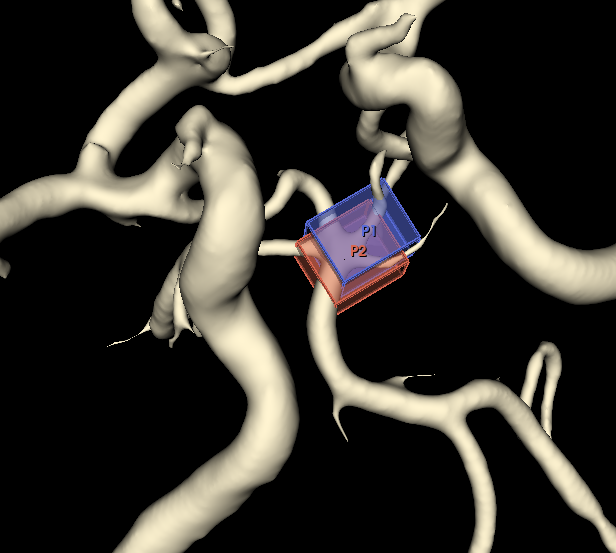}
\caption{3D crop of a segmented MRA-TOF image of dimension (445, 475, 252).
 Patches P1 and P2 with respective centers coordinates (226, 217, 147) and (224, 218, 151) overlap and encompass the same BoI. \label{CropsOverlapBifCtr}}
\end{figure}

Additionally, it is worth noting that for some peculiar anatomical configurations, some bifurcations may split into more than 2 daughters giving a trifurcation, a quadrifurcation or even more daughter branches (although uncommon), thus making the recognition step more challenging. 
Fig.~\ref{CropsOverlapBifCtr} depicts a portion of a 3D segmented MRA-TOF volume where 2 successive graph node neighbors are respectively the centers of two overlapping patches, both containing the same bifurcation K.
The patient-wise evaluation is performed using the automatic prior selection of patches. First, the raw image is resampled to $0.4 mm^3$, then an automated segmentation of the cerebral arteries is applied on the image using a 3D U-Net pretrained on our training dataset. Next, the 3D skeleton is computed on the obtained segmentation which allows the extraction of numerous patches centered at 3D graph nodes. The patches will then go through the classification network to estimate their label (refer to Fig.~\ref{GlobalScheme} to see the detailed flow chart). For each category of interest, we consider the most probable bifurcation (presenting the highest confidence score) based on the natural assumption that each category of BoI can only occur once for a particular patient. 
For a quantitative evaluation, we compute the distance between the predicted center of the bifurcation and the actual center of the ground truth BoIs annotated by the medical expert. 
For each inspected bifurcation, a recognition score of one is assigned if this distance falls below a given threshold. Otherwise, a value of zero is assigned. 
Any missing bifurcation, as specified by the ground truth labels, is not considered by the evaluation protocol. 
Overall, for each BoI, a recognition rate (recognized / existing bifurcations) can thus be computed on the test dataset.

%%%%%%%%%%%%%%%%%%%%%%%%%%%%%%%%%%%%%%
\section{Experimental Results}

Now that we have thoroughly described our image dataset, the neural networks and the evaluation protocol, let us present our experimental results.
We will first evaluate the global efficiency  on a full TOF acquisition, and then try to delve a bit deeper into the performances per bifurcation of interest.

\subsection{Segmentation results}
Let us first analyze the performances of the 3D U-Net onto the test set.
The 3D U-Net model achieved a Dice Similarity Coefficient of 0.84, a precision of  0.88 and a recall of 0.84 when segmenting the whole cerebral vascular tree. The 95HD value was 37 voxels. Such a result is close enough to the performances one can find in the literature with some reported models achieving a DSC up to 0.9 on MRA scans~\citep{wang2017,livne2019,chen2017,hilbert2020}.

\begin{figure}[!ht]
\begin{centering}
\includegraphics[width=0.9\columnwidth]{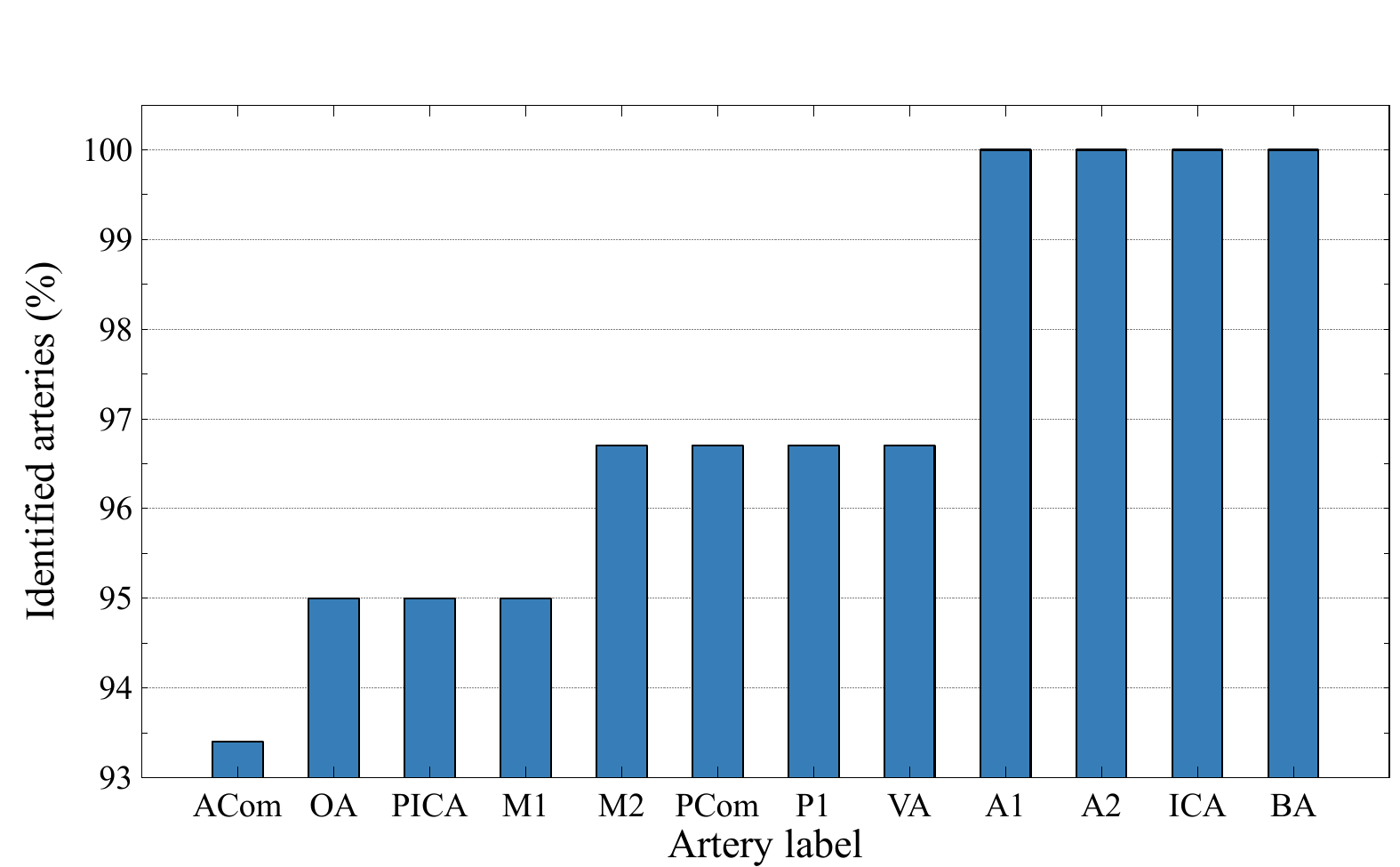}
\par\end{centering}
\caption{Percentage of identified arteries on the test dataset.\label{figseg}}
\end{figure}

The qualitative analysis concerning  the identification of the arteries constituting the Circle of Willis on the predicted U-Net segmentation masks is depicted in Fig.~\ref{figseg}. We report the percentage of arteries being properly identified (as described in Section \ref{sc}) for each artery composing the CoW. 
Overall, the larger arteries (\textit{BA}s, and \textit{ICA}s) were always correctly identified ($100\%$), whereas the smaller arteries (\textit{ACom}s, \textit{PComs}, \textit{OA}s and \textit{PCA-P1}s) were respectively identified with an accuracy of $93\%$, $96.7\%$, $95\%$, and $96.7\%$. 
Among the tested medium-sized arteries (\textit{MCA}s, \textit{PICA}s, \textit{VA}s and \textit{ACA}s), $97.2\%$ were correctly identified. In the rare cases of missed identifications, the annotators reported the presence of a gap in the segmented branch around the corresponding bifurcation.

%%%%%%%%%%%%%%%%%%%%%%%%
\subsection{CNN architecture benchmark}

To improve the classification performances, the 3D CNN design is of paramount importance, we have thus tested 4 different architectures and analyzed their performances in terms of BoIs classification. Very high cross-validation values are reached with all four architectures for labeling the bifurcations of interest, as shown in Table \ref{tab3}. 

\begin{table}[!ht]
\begin{center}
\caption{Performance measures reported on the cross-validation on various architectures.\label{tab3}}
\begin{tabular}{|c|c|c|c|c|}
\hline
\textbf{Model}&\textbf{Augs}&\textbf{ACC}&\textbf{AUC}&\textbf{F1-score}\\
\hline
VGG-16 & no augs & 86.45 & 97.13&84.76 \\
\hline
VGG-16 & flip & 89.80& 97.53 & 87.77\\
\hline
ResNet-18 & no augs &89.87 & 98.09&88.72\\
\hline
ResNet-18 & flip &91.08 & 98.89 & 89.71\\
\hline
DenseNet-121 & no augs &91.68 & 99.25 &90.31 \\
\hline
DenseNet-121 & flip &92.53 & 99.30 & 91.68\\
\hline
Proposed & no augs &92.71 & 99.14 &91.80 \\
\hline
Proposed & flip &\textbf{93.50} & \textbf{99.40} & \textbf{92.13}\\
\hline
\end{tabular}
\end{center}
\end{table}

However, our proposed Convolutional Neural Network (7M parameters) presents the best overall classification performances, as measured by all the metrics ($93.5\%$ ACC, $99.4\%$ AUC and $92.13\%$ F1-score), followed by DenseNet-121 (11M parameters), ResNet-18 (33M parameters) and VGG-16 (44M parameters). 
Our proposed architecture is thus preferred in the following, as it exhibits the best performances. 
Table~\ref{tab3} also shows that using axial flip as a way to augment the dataset may substantially improve the results.

%%%%%%%%%%%%%%%%%%%%%%%%
\subsection{Cross-validation classification results}

\begin{figure}[!ht]
\centering
\includegraphics[width=\columnwidth]{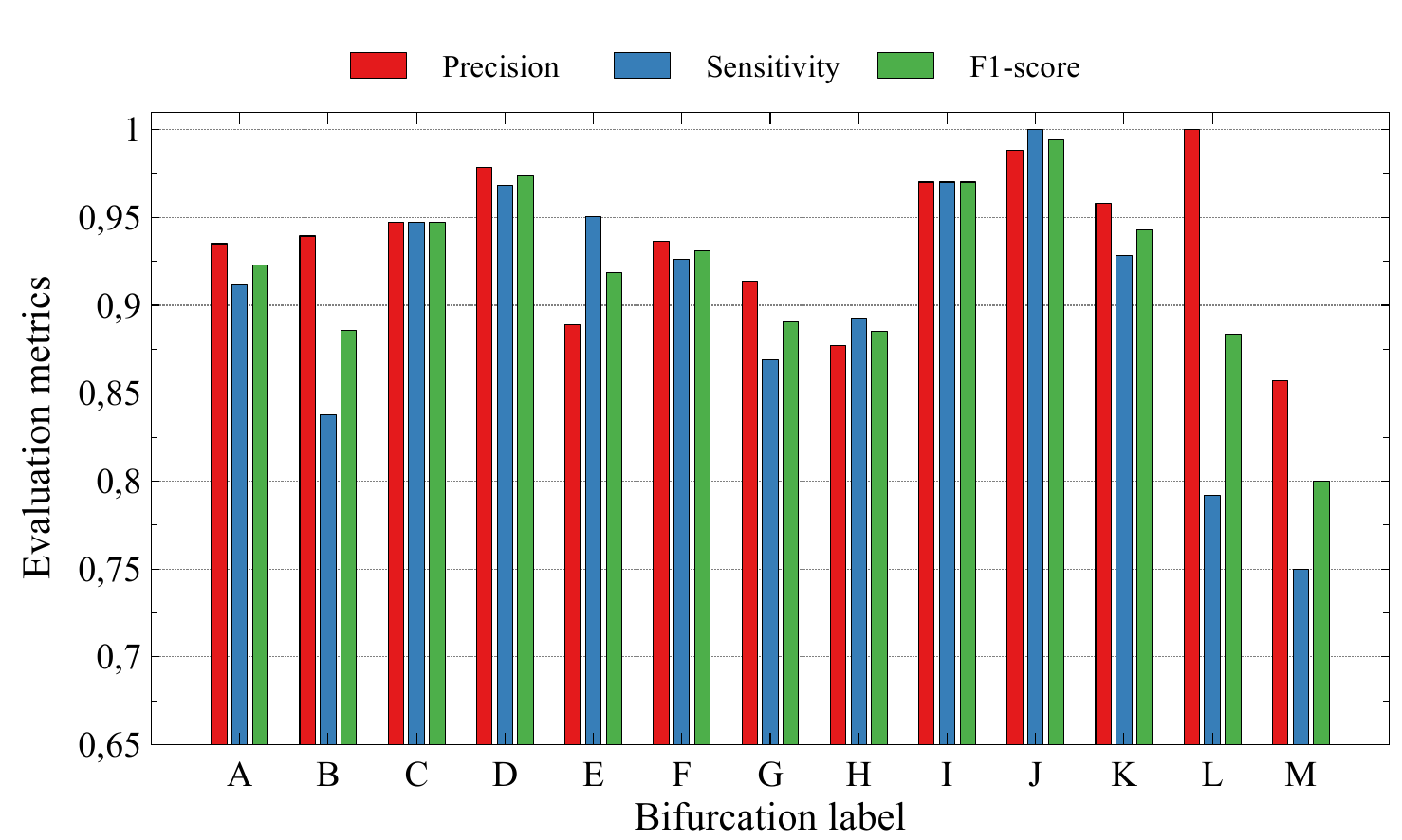}
\caption{ Classification performance of the 13 BoIs obtained from cross-validation. \label{DetectionRatePerBif}}
\end{figure}

For the classification experiments, our best model network discriminated between 13 classes with an overall  accuracy of 0.935 and an F1-score of 0.921. We report Precision, Sensitivity and F1-score for each the 13 bifurcations of interest in Fig \ref{DetectionRatePerBif}.

Overall, the model performs well for most classes, but with a slight performance decrease for bifurcations L and M. The best efficiency can be observed for classes C, D, I, J and K with an F1-score of 0.947, 0.973, 0.970, 0.994, 0.943 respectively. The sensitivity drops for L and M (0.792 and 0.750 respectively). 

Some L and M bifurcations are mistakenly classified as background bifurcations (BoNI, outside the CoW). This does not come as a surprise, as these bifurcations are missing in $80\%$ of our dataset due to an improper image acquisition. 
F1-score drops to $89\%$ for BoIs G and H. We observe that mislabeling mostly occurred in patches where the \textit{PCom} artery is either hypoplastic or even missing (aplasia). 
Other forms of miss-classifications occurred for the B bifurcation, which was occasionally detected as being A. 
More insights concerning the model's performances on various CoW configurations will be given in Section
\ref{disc}.

%%%%%%%%%%%%%%%%%%%%%%%%
\subsection{Quantitative results on the test set}

\begin{figure*}[t]
    \centering
  \subfloat[Using Expert segmentation \label{expert}]{%
       \includegraphics[width=0.44\linewidth]{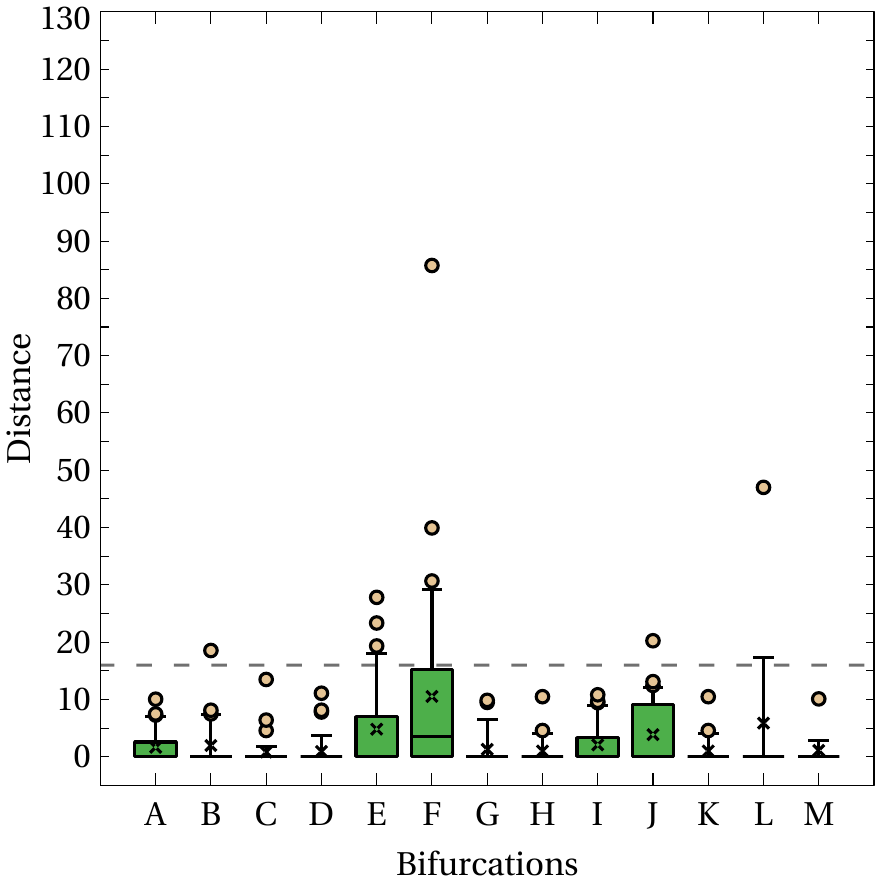}}
    \hfill
  \subfloat[Using U-Net segmentation \label{unet}]{%
        \includegraphics[width=0.44\linewidth]{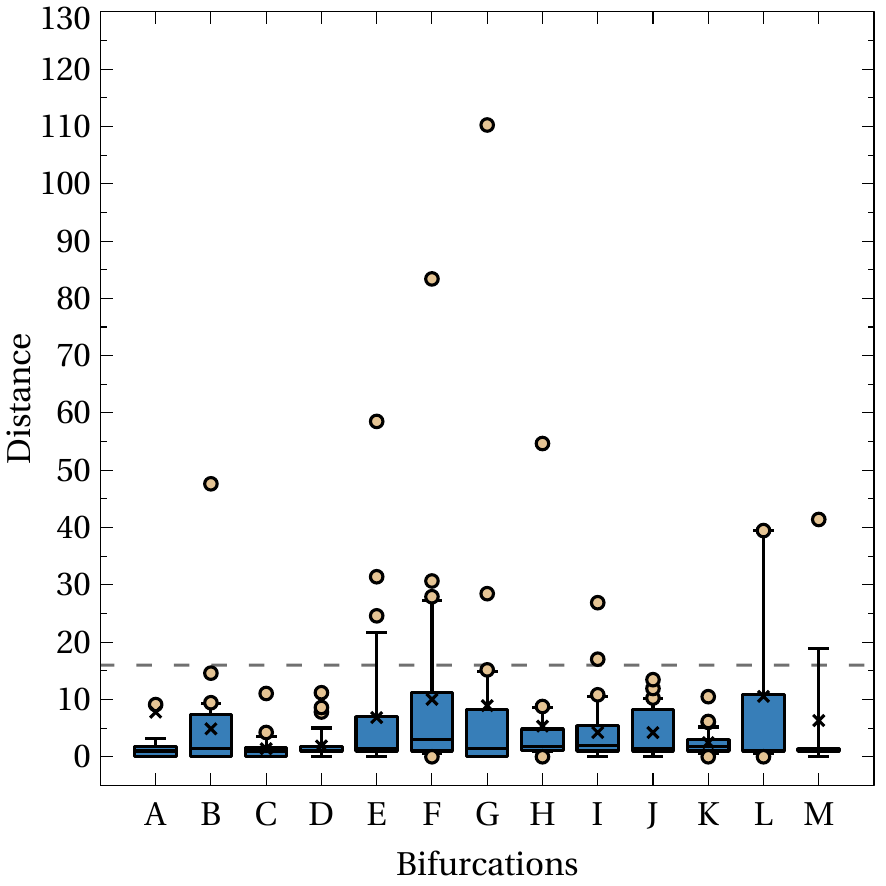}}
\caption{\label{BoxPlotsDistancePred} Distance distribution between the predicted bifurcations and the Ground Truth annotations on the test set.}
\end{figure*}
The test set composed of 30 MRA-TOFs, is independent from the training set. Its purpose is to evaluate the automated bifurcation recognition within a TOF acquisition. 
In fact, for a given TOF, all bifurcations issued by its 3D graph constitute the test set and will be labeled accordingly by the CNN. 
Each bifurcation either falls into one of the 13 BoI classes, or will be  assigned to the $14^{th}$ class (BoNI).

For quantitative results, we have conducted a comparison between the labels issued from the expert segmentation and the automatic U-Net segmentation. 

Fig.~\ref{BoxPlotsDistancePred} displays, for each category, the distribution of the 3D Euclidean distance between the predicted bifurcation center and the ground truth bifurcation center, using either the expert segmentation (a) or the U-Net segmentation (b). Considering the expert segmentation, across all subjects and all classes, $94\%$ of the predicted coordinates are located within a 16 voxels radius from  the annotation coordinates. 
Given that our 3D square patches are 32 voxels wide, setting a distance threshold at $Th=16$ voxels seems reasonable to consider that any predicted patch exhibiting a smaller distance will inevitably encompass the bifurcation of interest.

Examining the inter-quartile ranges shows a lower distance dispersion for bifurcations A, B, C, D, G, H and K. This can be attributed to their highly distinguishable geometry. 
Indeed, the shape of these bifurcations is quite unique, and hence, no other cerebral bifurcations do share a similar layout. 
However, we can observe in Fig.~\ref{BoxPlotsDistancePred}(a) that the distance distributions of bifurcations E and F are more spread out than any other bifurcation.

While Fig.~\ref{BoxPlotsDistancePred}(a) shows the distances from the predicted bifurcations to the ground truth labels on the manually segmented images, the labeling performances based on the U-Net segmentation are shown in Fig.~\ref{BoxPlotsDistancePred}(b). We can observe some rather strong similarities between both plots. 
Indeed, $91\%$ of the predicted patches are no further away than the defined distance threshold. 
More dispersed data was expected as the ground truth bifurcations annotations were based on the expert segmentation. However, ultimately, this had little effect on the overall recognition rate and BoI centers were localized closely to the coordinate annotations. 
Furthermore, some outliers can also be observed for bifurcations G and H when using the U-Net segmentation.

\begin{figure}[!ht]
\begin{centering}
\includegraphics[width=0.9\columnwidth]{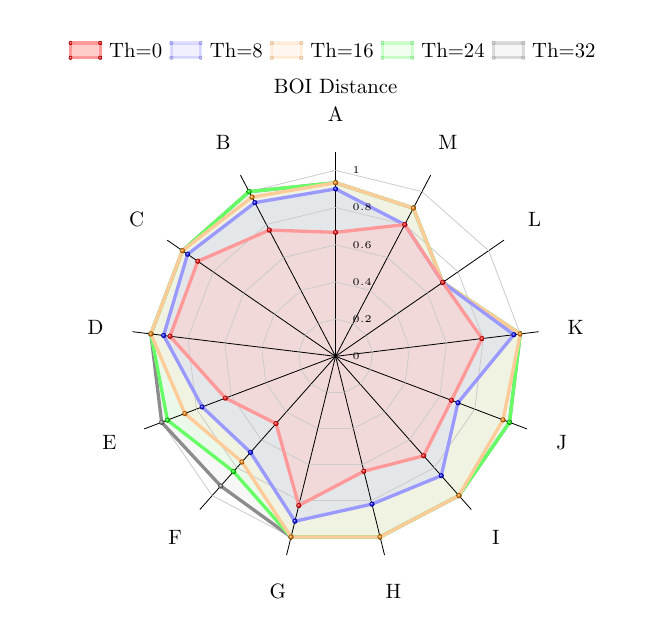}
\par\end{centering}
\caption{Recognition rates with respect to different distance thresholds. \label{RadarAccIncreasedTh}}
\end{figure}

We show in Fig.~\ref{RadarAccIncreasedTh} the evolution of the recognition rates for various distance thresholds. 
The overall recognition rate ranges from $73\%$ using a distance threshold of zero up to $97\%$ for a distance threshold set to the patch width ($Th=32$). The most significant performances decrease arises for bifurcations E and F. By adjusting the distance threshold, we observe a consistently higher recognition rate for bifurcations C, D, and K, while the recognition rate for L and M remains low.

\begin{figure}[!ht]
\begin{centering}
\includegraphics[width=1\columnwidth]{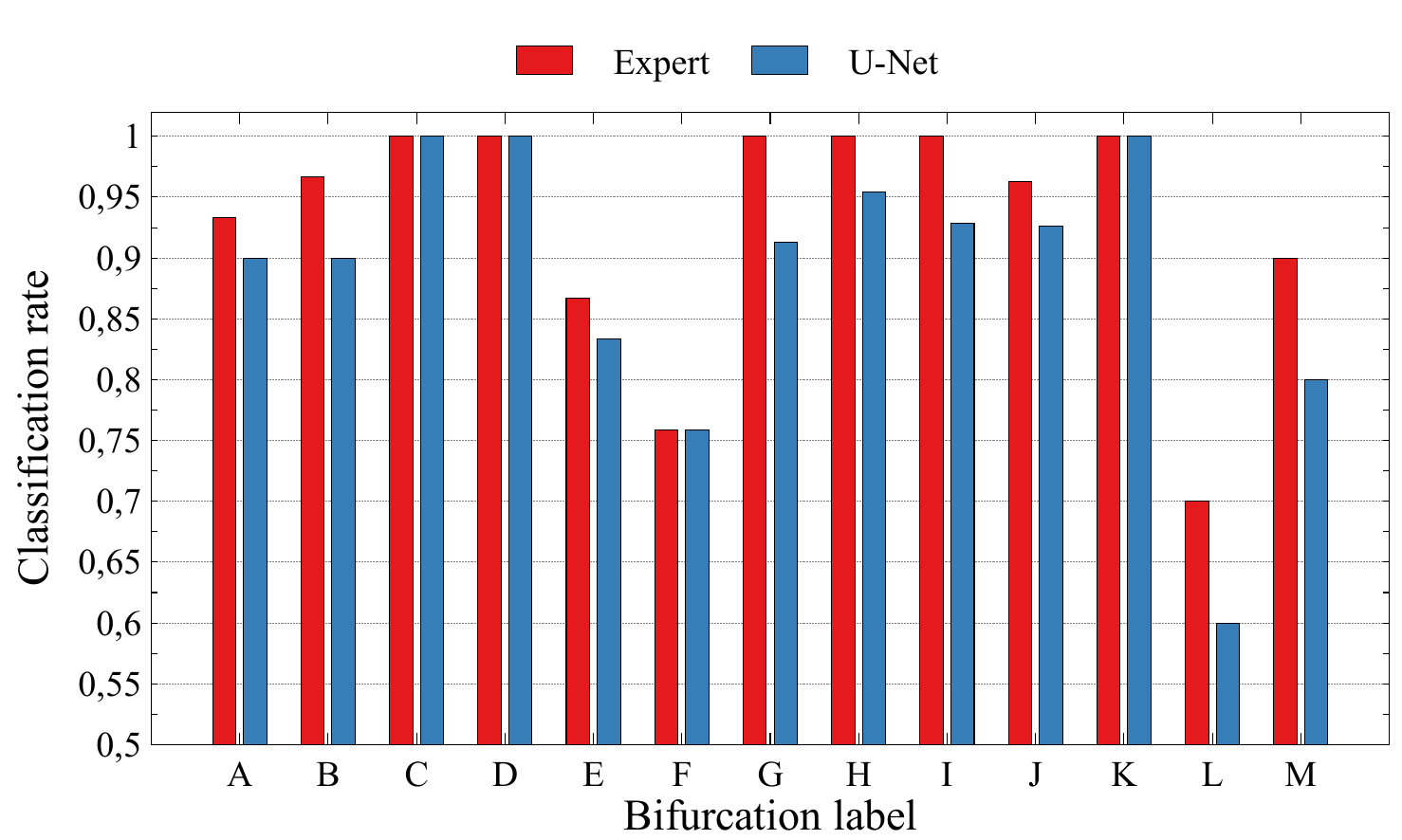}
\par\end{centering}
\caption{ Labeling performance by BoI class with respect to Expert and U-Net segmentation (with a distance threshold set to 16). \label{PerfPerBifTh16}}\end{figure}

Fig.~\ref{PerfPerBifTh16} presents, for each BoI, the classification rate (with a $Th=16$ voxels) when the segmentation was either performed manually, or via the U-Net model. We can observe that, overall, the segmentation method has a very limited impact on the bifurcation recognition. In fact, the few encountered missed detection concern bifurcations with small  and/or thin branches : extremely small \textit{ACom} artery for A and B labeling, thin ophthalmic artery for I and J labeling, small \textit{PICA} artery for L and M labeling or hypoplastic/missing \textit{PCom} artery for G and H labeling.

\begin{figure*}[!ht]
    \begin{center}
        \begin{subfigure}{0.49\linewidth}
            \centering
            \includegraphics[height=0.63\linewidth]{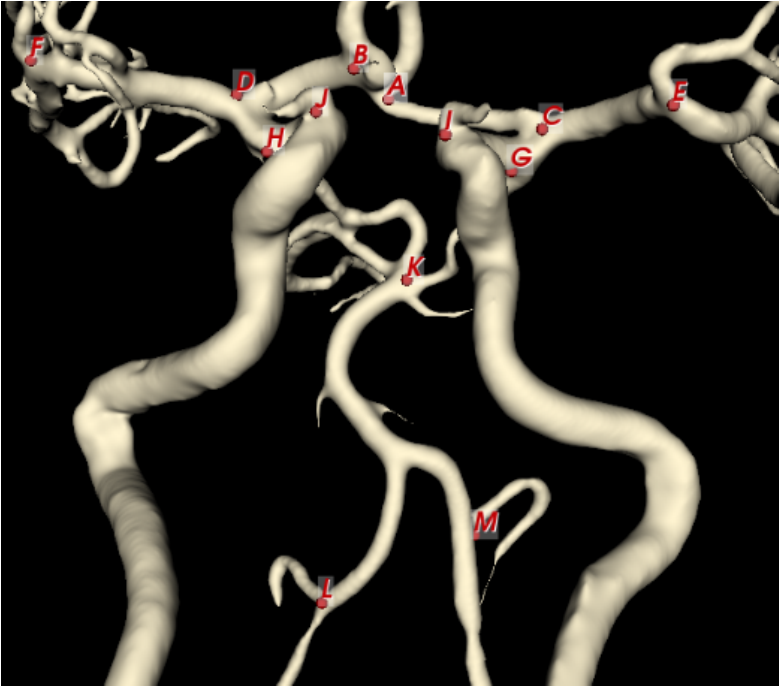}
        \caption{\label{3DExples_a}Example of an MRA-TOF with a complete Circle of Willis (all  BoI being correctly labeled).}
        \end{subfigure}
        \begin{subfigure}{0.49\linewidth}
            \centering
            \includegraphics[height=0.63\linewidth]{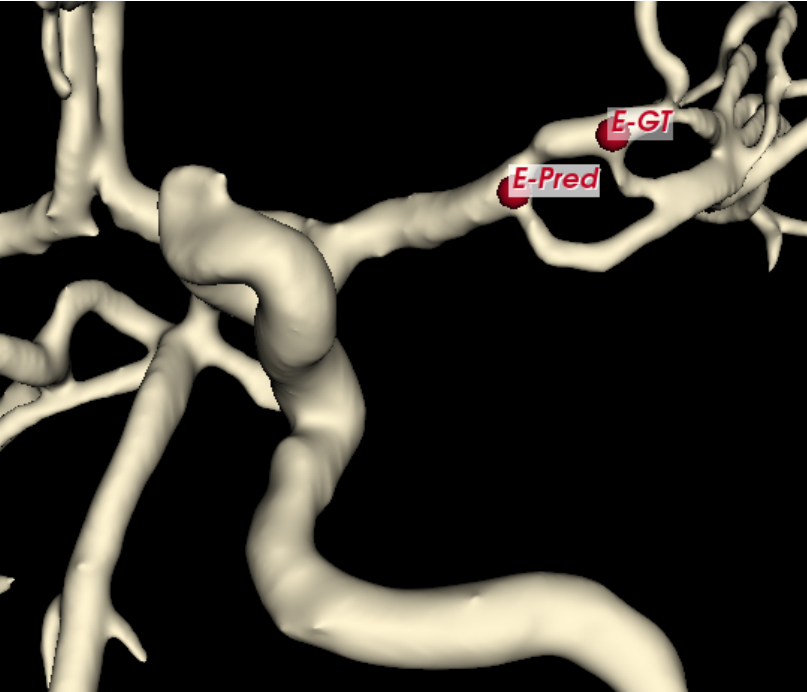}
        \caption{\label{3DExples_b}Incorrectly labeled Bifurcation E, detected 21 voxels away from the ground truth.}
        \end{subfigure}
        \begin{subfigure}{0.74\linewidth}
            \centering
            \includegraphics[width=0.74\linewidth]{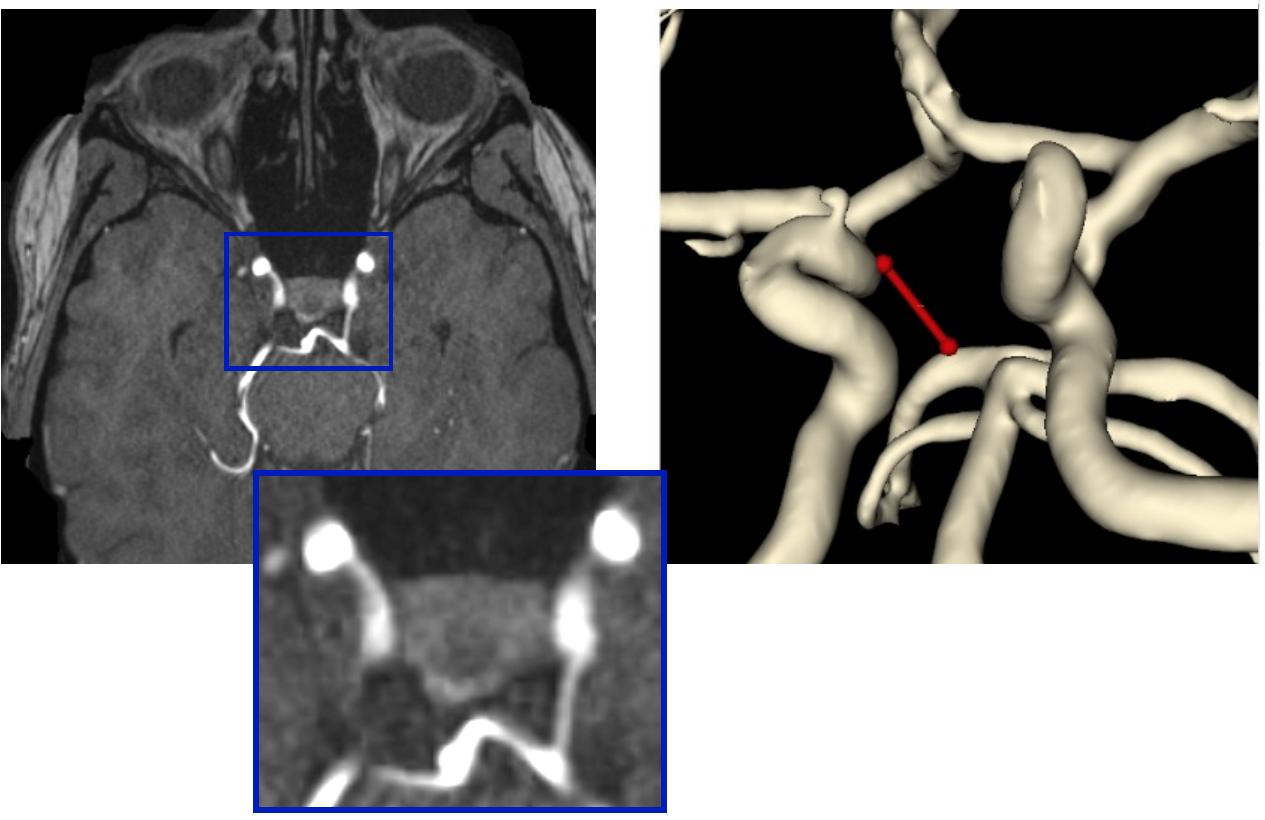}
        \caption{\label{3DExples_c}Right \textit{PCom} missing in the segmentation (partially hypoplastic), leading to an undetected H bifurcation.}
        \end{subfigure}
    \end{center}
    \caption{\label{3DVascuExamples}Qualitative analysis of predictions. For visualization purposes, some arteries outside the CoW were cut from the segmentation maps.}
\end{figure*}

%%%%%%%%%%%%%%%%%%%%%%%%%%%%%%%%%%%%%%
\section{Discussion and Conclusion}\label{disc}

In this work, we have proposed a method to provide an automatic anatomical labeling of the bifurcations constituting the Circle of Willis on human MRA-TOF acquisitions using Convolutional Neural Networks. Our approach combines an efficient segmentation of the cerebral arteries and a patch-wise classification of BoIs using a 3D CNN. 
From a training set of pre-annotated patches, the model was able to classify the thirteen bifurcations of interest with an accuracy score of $93.5\%$.
Indeed, for an automatic BoI detection and labeling (recognize a given BoI within an entire MRA-TOF image), we have tested our method on an independent test set composed of 30 MRA-TOFs. On this dataset, $91\%$ of the bifurcations were correctly identified within a 16 voxels distance radius using the U-Net pre-segmentation step. These results are highly encouraging, as they prove that the model is capable of isolating and labeling the main bifurcations of the Circle of Willis without any user intervention and can be applied to new images never seen before by the model. 

Overall, the BoI were detected and labeled with a  high rate ($>90\%$) for 9 of the 13 categories, and even up to $100\%$ for bifurcations C, D and K.
The labeling rate for L ($60 \%$) and M ($80 \%$) is significantly reduced, compared to the overall recognition rate. Such a reduced efficiency was expected, and can be explained by a more limited representation of these bifurcations in our image dataset ($20 \%$ of MRA-TOFs). Indeed, as Fig.~\ref{3DVascuExamples}(a) shows, such bifurcations are a bit remote from the rest of the CoW, and quite frequently the MRI technician starts the acquisition above L and M. Labeling errors also occurred for the bifurcations E and F, due to the geometry variability of the \textit{MCA} branch. It is widely accepted in the literature, that the bifurcations E and F along the \textit{MCA} exhibit a rather complex geometry \citep{tanriover2003} and, most importantly, there exists a significant patient variability leading the neuroradiologists community to disagree on the exact location of the \textit{MCA-M1} / \textit{MCA-M2} splitting~\citep{rhoton2002,zaidat2013,findakly2020}, and hence on the precise location of bifurcations E and F. A visual representation of the segmentation map corresponding to an incorrect E labeling is shown in Fig.~\ref{3DVascuExamples}(b). The location of the predicted E bifurcation fell 21 voxels away from its corresponding ground truth.

These results are very promising as the tests were performed on a large panel of MRA acquisitions and device types. The comparison with other related works \citep{Robben2016, bogunovic2013} is difficult, as these latter are not fully automatic and require a prior manual selection of BoIs by experts, which is time consuming. Dumais et al. \citep{DUMAIS2022}, have recently developed the only automated method devoted to the CoW arteries segmentation. However, the authors only considered a limited portion of the CoW. For instance, only the first segment of the \textit{MCA} was accounted for, and moreover, bifurcations E, F, I, J, L, M, presenting a high risk of aneurysm occurrence ($27 \%$, $26 \%$ and $8\%$) were \textit{de facto} discarded from the automatically cropped area. 

Although this work is not focused on the segmentation performances, we have shown that our U-Net model may yield a slightly lower Dice coefficient compared to some other vasculature segmentation from the literature. However, all competing projects have been conducted on an image dataset acquired using either one or two distinct scanners. In contrast, our study involved 12 different MRI scanners, which may account for the slightly lower Dice score observed in our dataset. 

Regarding the segmentation pre-processing step, it is worth noting that, while achieving a high Dice coefficient is desirable, it is not the most critical factor. Of paramount concern, is the accurate representation of the CoW bifurcations within the 3D graph. Specifically, all arteries forming the bifurcations of interest along the Circle of Willis should be faithfully segmented, in order to ensure that the vicinity of each researched bifurcation will be correctly located within the 3D graph. In other words, the segmentation step serves as a pre-selection process, locating all bifurcations that will be subsequently classified by the 3D CNN. Hence, ultimately, the segmentation accuracy is not crucial, but it is imperative that the segmented images (or more precisely their 3D skeletons) encompass all arteries composing the CoW bifurcations. A missing segmented artery would inevitably lead to missed bifurcations that will not be explored by the 3D CNN during the bifurcation classification process. Our study compares expert segmentation with U-Net segmentation and shows that the recognition rate is marginally affected by the segmentation performance.

Compared to the expert segmentation, the recognition rate using the U-Net pre-segmentation suffers a $3\%$ decrease (from $94$ to $91\%$). Bifurcations that involve larger and medium-sized arteries are correctly identified. Miss-detected bifurcations were mostly due to non-segmented \textit{AComs} or \textit{OA} arteries (extremely thin), and unsegmented \textit{PComs} arteries. In most cases, the missed detections concern partially hypoplastic branches that could be confused with noise (Fig.~\ref{3DVascuExamples}(c)). Moreover, we could explain a missing artery by the anatomical variability of the CoW rather than the segmentation process. Indeed, compared to ground truth segmentation, only $3.3\%$ of the \textit{PComs},
$6.7\%$ of the \textit{AComs} and $5\%$ of the \textit{OAs} were not  segmented by the U-Net.  This suggests that the current work can nevertheless provide an excellent bifurcation automatic recognition.
Lastly, possible improvements of our model mostly include a better representativity of the labeled L and M patches in the dataset. It is worth noting that, in the literature, no other works consider these two bifurcations as part of the CoW; Nevertheless, in this study, we have decided to consider these two bifurcations as the risk of developing an aneurysm there is somewhat significant.  To improve performance and tackle the class imbalance problem, we need to explore alternative techniques such data augmentation, which involves increasing the diversity and size of the training dataset. However, traditional data augmentation methods may not be directly applicable to our medical image analysis task.
Interestingly, the only augmentation technique that had a positive impact on accuracy in our study was horizontal flipping. It is clear that developing specialized augmentation techniques tailored to our task is necessary to effectively address the class imbalance problem.
We might also conceive a sub-division of bifurcations E and F into two distinct classes, given the relative uncertainty encountered by the neuroradiologists on the \textit{MCA} branch geometry.

%%%%%%%%%%%%%%%%%%%%%%%%%%%%%%%%%%%%%%
\section*{acknowledgements}
This work was partially supported by the French ANR project ``WECAN'' (ANR-21-CE17-0006).

\bibliographystyle{model2-names.bst}\biboptions{authoryear}
\bibliography{refs}

\end{document}